\documentstyle[12pt]{article}
\def \as {\alpha_s}
\def \gl {\tilde{g}}
\def \ms {{\overline{\mbox{MS}}}}
\def \dr {{\overline{\mbox{DR}}}}
\newcommand{\prepr}[1] {\begin{flushright} {\bf #1} \end{flushright} \vskip
1.5cm}
\newcommand{\titul}[1] {\begin{center}{\large\bf #1 } \end{center}\vskip 1.cm}
\newcommand{\autor}[1] {\begin {center} {\large \lineskip .5em #1 }
                        \end   {center} }
\newcommand{\lugar}[1] {\begin{center} {\it #1} \end{center}}
\newcommand{\abstr}[1] {{\begin{center} \vskip .5cm {\bf Abstract
                        \vspace{0pt}} \end{center}}\begin{quote} #1
                        \end{quote}}
\begin{document}

\begin{titlepage}
\prepr{UCI-TR/94-3\\ January 1994}
\titul{Light gluinos and the longitudinal structure function}
\autor{A.V. Kotikov}
\lugar{Laboratory of Particle Physics, Joint Institute for Nuclear Research\\
101 000 Dubna, Russia}
\autor{G. Parente}
\lugar{Department of Physics, University of California  \\
Irvine, CA 92717, USA}
\autor{O.A. Sampayo}
\lugar{Departamento de F\'\i sica,\\
CINVESTAV-IPN, Mexico }
\abstr{ The leading effect of light gluinos on the
deep inelastic longitudinal
structure function is calculated. We present
the explicit analitic expression
for the Wilson coefficient.
After convolution with quark, gluon and gluino distributions
we found that the size of the contribution is of order a
few percent of the total $F_L$. Some
phenomenological implications for HERA and LEP/LHC are given.}
\end{titlepage}
\newpage

The modification of the standard QCD
predictions due to the existence of supersymetric particles
has been extensively investigated in the last decade. It was suggested
that the effect of some of these particles could be detected in the
(low) energy range of current experiments. In particular a light gluino,
a strongly interacting neutral fermion which is the spin-1/2 partner of the
gluon, of mass of a few GeV, has not been ruled
out by experiment\footnote{Recently it was claimed \cite{NANOPOULOS}
that the light gluino scenario is disfavoured by current LEP data on
chargino and Higgs assuming a radiative symmetry breaking of the
gauge group but there is controversy on that point \cite{VALLE} }
and even its effect (if they exist) could explain a certain
claimed discrepancy between low and high energy measurements
of the strong coupling constant (see for example \cite{STIRLING}).

The effect of gluinos on the scaling violation of deep inelastic
structure functions has been calculated in perturbative QCD at
leading order (LO) \cite{GLUINOSLO} and at next-to-leading order (NLO)
\cite{ANTONIADIS}. As the contribution was found to be smaller than
the precision of the data, it can be concluded that such
experiments can not exclude the
presence of gluinos of mass lower than 10 GeV.
Recently the gluino effect on $F_2$ was investigated in the kinematical
range of HERA using modern parton distributions at NLO and it was found
to be small and of the size comparable to the effect due to the
uncertainty in the QCD energy scale, which makes it very hard
to detect.

Until now the part of the interaction
represented by the Wilson coefficient,
where the gluino starts to contribute at order $\as^2$, was never considered
in the calculations. Although for $F_2$ it is expected to be small
because the correction follows in importance to those from quarks and gluons
proportional to $\as^0$ and $\as^1$,
for $F_L$ which starts at order $\as$ the effect could be relevant and an
explicit calculation is required.

The aim of this work is to investigate the effect of light gluinos
on the longitudinal structure function and to explore the posibility
of detection in HERA or larger colliders.
Now the calculation can be easily done because one has
enough information about the two
building blocks in a parton model approach, i.e. the hard
part of the interaction and the gluino momentum distribution
inside the proton.

We have translated to the gluino case the results
from some of the quarks and gluon diagrams contributing to
the O($\as^2$) $F_L$  Wilson coefficient.
For the soft part of the interaction we use
a recent set of parton densities determined
from global fits which includes the influence of a light gluino
\cite{STIRLING}.
Throughout this work we will assume the hypothesis used in the
extraction of such
parton and gluino distributions, i.e. that the gluino is of Majorana type
of mass of 5 GeV
\footnote{ In our calculation the value of the mass defines
the threshold point, $Q^2=4 m_{\gl}^2$, above which the gluino
evolves as a massless parton, and below it is zero. It also affects
the running coupling constant through the coefficients of the
$\beta$-function (see ref. \cite{STIRLING} for details). An alternative
mass dependent $\as$ evolution can be found in ref. \cite{SHIRKOV}}.
We do not consider the influence of
the scalar partners of quarks because it is assumed they are heavy
enough to be decoupled in the low energy region.

The diagrams in fig. 1 are the Compton amplitudes at the lowest
order involving gluinos.
Their contribution can be extracted from the analogous quark
diagrams, those with the gluino line being replaced by a quark line,
with an appropriate change of the group factors.
The result of diagram (a) can be read from the term proportional
to $C_F n_f$ in the non singlet coefficient (see eq. (8) from \cite{NPB}),
changing the number of species in the quark loop ($n_f$)
to $C_A$, which is the corresponding number in case
of a Majorana-type gluino.
Also, the contribution of diagrams in fig. (b) can be obtained from the
singlet coefficient ( eq.(9) of the same reference) with the
substitution $C_F \rightarrow C_A$.

The gluino contribution to $F_L$ is connected to the quark and gluino
($\gl$)
densities by the parton model relation (see \cite{NPB} for notation):
\begin{eqnarray}
            F_L^{\gl} (x,{ Q ^2})& = & \left( \frac{\as}{4 \pi} \right)^2
            \int _x ^1 \frac{dy}{y} f_{L,q}^{\gl(2)} (y)
            {\cal F} ( x / y ,{ Q ^2})
\nonumber \\ [ .4 cm]
& & \mbox{} + \left( \frac{\as}{4 \pi} \right )^2
 \int _x ^1 \frac{dy}{y}  f_{L,\gl}^{(2)} (y)
    \delta _\psi ^2
    {\cal G}_{\gl} ( x / y, \,  Q ^2) \,,
                                                   \label{eq:flgluino}
\end{eqnarray}
where
\begin{eqnarray}
{\cal F}   (x, \, { Q ^2}) & = &
    \sum _{i=1}^{n_f} e_i^2 x
    \left( q_i (x, \, { Q ^2})
    + \bar{q_i}(x, \, { Q ^2}) \right) ,
\nonumber \\ [ .4 cm]
{\cal G}_{\gl}   (x, \, { Q ^2}) & = &
     x\gl (x, \, { Q ^2}) ,
\,\,\,\,\,\, \delta _\psi ^2 =
    \left( \frac{ \sum_{i=1}^{n_f} {e_i}^2}{n_f} \right)
\label{eq:QUS}
\end{eqnarray}
The coefficients in eq. (\ref{eq:flgluino}) that were calculated with the
help of the changes mentioned above have the following
form in the $\ms$ scheme:
\begin{eqnarray}
\lefteqn {f_{L,q}^{\gl(2)}(x) =
    - \frac {8}{3} C_F C_A x^2 \left[ \ln \left(
      \frac {x^2}{1-x}\right) - \frac {6-25x}{6x}
    \right] } \,,
\label{eq:kerns}
\\ [ .6 cm]
\lefteqn {f_{L,\gl}^{(2)}(x) =
    \frac {16}{9} C_A n_f
    \left[
      3 \left( 1-2x-2x^2 \right)
      (1-x)\ln(1-x)
    \right.}
\nonumber \\ [ .4 cm]
& & \mbox{} + 9x^2
         \left( Li_2 ( x \right) + \ln^2 (x) - \zeta(2) )
      + 9x (1 - x - 2x^2)\ln x
\nonumber \\ [ .4 cm]
& & \left.
      \mbox {}
     - 9x^2 (1-x) - (1-x)^{3}
    \right] \, ,
\label{eq:kers}
\end{eqnarray}
Subindexes q and $\gl$ label
the type of parton distribution in the convolution integral.
$n_f$ is the number of active quarks flavours. $C_F$ and $C_A$ are
the Casimir operators in QCD. $\zeta$ is the Riemann
zeta-function and $Li_2$ is the dilogarithm function.

Although in our calculation we use the $\ms$ results given above in
order to be (scheme)consistent with the parton densities,
for completeness we present below the transformations
to a renormalization scheme which preserves supersymmetry and gauge invariance
and that has been extensively used in the literature, i.e. the so-called
dimensional reduction renormalization scheme $\dr$ \cite{DRSCHEME}.
The relation between quark and gluon coefficients in both schemes are:
\begin{eqnarray}
f _{L, \, q} ^{NS(2)}(\dr) & = & f _{L, \, q} ^{NS(2)}(\ms)  -
 f _{L, \, q} ^{(1)} \otimes \Delta O_{qq}
\nonumber \\
f _{L, \, q} ^{S(2)}(\dr) & = & f _{L, \, q} ^{S(2)}(\ms)  -
 f _{L, \, G} ^{(1)} \otimes \Delta O_{Gq}
\label{eq:conect1} \\
f _{L, \, G} ^{(2)}(\dr) & = & f _{L, \, G} ^{(2)}(\ms)  -
 f _{L, \, q} ^{(1)} \otimes \Delta O_{qG}
 - f _{L, \, G} ^{(1)} \otimes \Delta O_{GG}
\nonumber
\end{eqnarray}
$f _{L, \, q} ^{NS(2)}, f _{L, \, q} ^{S(2)}$ and $f _{L, \, G} ^{(2)}$
in  $\ms$ can be found, for example, in ref. \cite{NPB} (with the
corrections given in \cite{CORRECTO} or \cite{PRLMULTIP}
for the gluon coefficient).
$f _{L, q(G)}^{(1)}$ are the first order coefficients.
$\otimes$ is the convolution symbol.
For the pure gluino contributions, Eq.
\ref{eq:kerns} and \ref{eq:kers}, it reads:
\begin{eqnarray}
f _{L, \, q} ^{\gl(2)}(\dr) &=& f _{L, \, q} ^{\gl(2)}(\ms)
\nonumber \\
f _{L, \, \gl} ^{(2)}(\dr) &=& f _{L, \, \gl} ^{(2)}(\ms)  -
 f _{L, \, G} ^{(1)} \otimes \Delta O_{Gq}
\label{eq:conect2}
\end{eqnarray}
The moments of the functions $\Delta O_{ij}$, $i,j=q,G$
(see \cite{ANTONIADIS} for notation) are the pieces that must
be added to the matrix elements of the Wilson operator
in the $\ms$
scheme to get the result in $\dr$ \cite{ANTONIFLORATOS}.
We have calculated the convolutions of eq. (\ref{eq:conect1})
obtaining:
\begin{eqnarray}
 f _{L, \, q} ^{(1)} \otimes \Delta O_{qq} &=&
  8 C_f^2 x ( 1 - \frac{3}{2} x + x \ln x )
\nonumber \\
 f _{L, \, q} ^{(1)} \otimes \Delta O_{qG} &=&
  0
\nonumber \\
 f _{L, \, G} ^{(1)} \otimes \Delta O_{Gq} &=&
  16 n_f C_f x^2 ( 1 - x + \ln x )
\label{eq:cambioDR} \\
 f _{L, \, G} ^{(1)} \otimes \Delta O_{GG} &=&
- 32 n_f C_A x^2 \left( (1+x)\ln x + \frac{25}{12}(1-x) \right)
\nonumber
\end{eqnarray}
The coefficients from diagrams involving gluinos
are shown in fig. 2 for the $\ms$ case and $n_f=4$.
For comparison the well-known contributions
from quarks and gluon are plotted.
We checked that the use of the $\dr$ scheme does not change
significantly its value and shape.

To calculate the longitudinal structure functions
Eq. (\ref{eq:flgluino}) was integrated numerically convoluting the above
coefficients with parton densities from set MRS-D$_0'$ and
also from the version with gluinos from ref. \cite{STIRLING}.
We have considered the running coupling constant at NLO and
the gluino threshold was crossed with the same prescription used
for quarks, i.e.
matching $\as$ above and below threshold changing the QCD
energy scale $\Lambda$ to compensate the change in the
renormalization group $\beta$ function.

The result is presented in
fig. 3, where for a better appreciation of the gluino effect we
plotted the ratio $F_L^{\gl} (x,{ Q ^2})/F_L$, $F_L$ being the
standard quark and gluon contribution up to second order in QCD.
The difference is a few percent in the relevant kinematic
range for HERA and LEP/LHC. Notice that for $x$ values higher than 0.1 the
contribution becomes relatively more important, but there $F_L$ is very
small and will not be measured.

Whether it will be possible or not to detect the effect of gluinos
in $F_L$ is illustrated by the ratio plotted in figs. 4(a) and (b).
One can see that both predictions are very similar in almost the whole
region accesible
to HERA ($Q^2<10^5$ GeV$^2$, $10^{-5}Q^2<x<1$) and future
LEP/LHC ($Q^2<1.6\times10^6$ GeV$^2$, $6.25\times10^{-7}Q^2<x<1$)\footnote{
As we are considering the gluino as a massless particle
above the threshold,
the use of Eq. \ref{eq:kerns} (see diagram 1a) for $x$ and $Q^2$
close to the gluino production threshold is not correct.
A different approach to cross the discontinuity at the threshold similar
to those used for heavy quarks (see ref. \cite{SHIRKOV})
is needed but here we do not treat this problem}.

Notice how at low $x$ the ratio of structure functions rises
with $Q^2$ with the same shape as the
ratio of coupling constants, which is what one naively would expect
from $F_L$ being proportional to $\as$.
At higher $x$ this effect is not so pronounced due to the balance
from others contributing terms.
A similar behavior can be observed in $F_2$ (see fig. 2 in ref.
\cite{STIRLING}), i.e. at low $x$ the structure function is
dominated by sea-quarks and gluons that in turn depend on $\as$.

Similar results as those obtained above
for the size and shape of the gluino contribution
hold for the
ratio of the longitudinal and transverse
cross-sections $R=\sigma_L/\sigma_T=F_L/2xF_1$.

In conclusion, the influence of light gluinos in $F_L$ is of order
of a few per cent in the $x$ and $Q^2$ range of current and
future machines. For example, at $Q^2=400$ GeV$^2$ and $x=10^{-2}$
the gluino would produce a 2 $\%$ negative contribution. In LEP/LHC
and for the same $Q^2$ one could go down to $x=10^{-3}$ and see as
much as a 3.5 $\%$ effect.
These values are smaller than the
contribution of the second order
QCD correction for a non singular gluon but
similar in size to the correction from a steep gluon of type
$x^{-3/2}$ \cite{PRLMULTIP}.
Only a very precise measurement of $F_L$, which is not expected
at HERA, could separate both second order gluon and gluino
effects.

\vskip 0.5 cm
%
%
%------------   FIN -------------
%
%
\noindent{\bf Acknowledgments }

We are grateful to J.W. Stirling for providing
the parton distributions used in this work.
One of the authors, Gonzalo Parente,
acknowledges the financial
support from the `Comision Interministerial de Ciencia y
Tecnolog\'\i a', Spain.
%
%                 REFERENCIAS
%

%
%
\newpage
\noindent{\bf Figure captions }
\vspace{.5cm}
%
%
%
%
%\begin{figure}
%
%\caption{

\vspace{1.cm}

\noindent{Figure 1:
The diagrams involving gluinos (dashed line) contributing
to the $F_L$ O($\as^2$) Compton amplitude.}

%
%\end{figure}
%
%\begin{figure}
%
%
%\caption{

\vspace{1.cm}

\noindent{Figure 2:
O($\as^2$) part of the $F_L$ quark, gluon and gluino
coefficients:
(a) $f_{L,q}^{NS(2)}$,
(b) $f_{L,G}^{(2)}\delta _\psi ^2$,
(c) $f_{L,q}^{S(2)}$,
(d) $f_{L,\gl}^{(2)}$ and
(e) $f_{L,q}^{\gl(2)}$. }

%
%\end{figure}
%
%\begin{figure}
%
%
%\caption{

\vspace{1.cm}

\noindent{Figure 3:
The solid lines are the gluino contribution to $F_L$ relative
to the NLO prediction at $Q^2=10^3$ and
$10^4$(lowest solid line at low x) GeV$^2$.
The dashed(dash-dotted) line is the contribution due to the interaction
represented by the diagrams in fig. 1a(1b) at $Q^2=10^3$ GeV$^2$.}

%
%\end{figure}
%
%\begin{figure}
%
%
%\caption{

\vspace{1.cm}

\noindent{Figure 4:
Ratio of structure functions $F_L$ calculated with and without
the influence of gluinos at fixed $Q^2$ (a) and $x$ (b). In (a)
the upper(lower) curve was obtained at $Q^2=10^4$($10^3$) GeV$^2$.
The dashed line in (b) is the evolution of the ratio of the corresponding
strong coupling constants.}

%
%\end{figure}
%
\end{document}